\documentclass[prl,twocolumn,longbibliography,superscriptaddress]{revtex4-1}

\usepackage{graphicx}
\usepackage{amsmath}
\usepackage{multirow} 
\usepackage{physics} 
\usepackage{braket}
\usepackage{subfigure}   
\usepackage{verbatim}
\usepackage{color}
\usepackage{tikz}

\newcommand{\el}[1]{\textcolor{black}{#1}}
\newcommand{\er}[1]{\textcolor{black}{#1}}
\newcommand{\erl}[1]{\textcolor{black}{#1}}

\begin{document}

\title{An energy efficient quantum-enhanced machine}

\author{Waner Hou}
\affiliation{CAS Key Laboratory of Microscale Magnetic Resonance and School of Physical Sciences, University of Science and Technology of China, Hefei 230026, China}
\affiliation{Anhui Province Key Laboratory of Scientific Instrument Development and Application, University of Science and Technology of China, Hefei 230026, China}
\author{Xingyu Zhao}  
\affiliation{CAS Key Laboratory of Microscale Magnetic Resonance and School of Physical Sciences, University of Science and Technology of China, Hefei 230026, China}
\affiliation{Anhui Province Key Laboratory of Scientific Instrument Development and Application, University of Science and Technology of China, Hefei 230026, China}
\affiliation{Hefei National Laboratory, University of Science and Technology of China, Hefei 230088, China}
\author{Kamran Rehan}
\email{Corresponding author; Email:  krehan2010@yahoo.com}
\affiliation{CAS Key Laboratory of Microscale Magnetic Resonance and School of Physical Sciences, University of Science and Technology of China, Hefei 230026, China}
\affiliation{Department of Physics, The University of Haripur, KP Pakistan}
\author{Yi Li}
\affiliation{CAS Key Laboratory of Microscale Magnetic Resonance and School of Physical Sciences, University of Science and Technology of China, Hefei 230026, China}
\affiliation{Anhui Province Key Laboratory of Scientific Instrument Development and Application, University of Science and Technology of China, Hefei 230026, China}
\author{Yue Li}
\affiliation{CAS Key Laboratory of Microscale Magnetic Resonance and School of Physical Sciences, University of Science and Technology of China, Hefei 230026, China}
\affiliation{Anhui Province Key Laboratory of Scientific Instrument Development and Application, University of Science and Technology of China, Hefei 230026, China}
\author{Eric Lutz}
\email{Corresponding author; Email:  eric.lutz@itp1.uni-stuttgart.de}
\affiliation{Institute for Theoretical Physics I, University of Stuttgart, D-70550 Stuttgart, Germany}
\author{Yiheng Lin}
\email{Corresponding author; Email:  yiheng@ustc.edu.cn}
\affiliation{CAS Key Laboratory of Microscale Magnetic Resonance and School of Physical Sciences, University of Science and Technology of China, Hefei 230026, China}
\affiliation{Anhui Province Key Laboratory of Scientific Instrument Development and Application, University of Science and Technology of China, Hefei 230026, China}
\affiliation{Hefei National Laboratory, University of Science and Technology of China, Hefei 230088, China}
\author{Jiangfeng Du}
\email{Corresponding author; Email:  djf@ustc.edu.cn}
\affiliation{CAS Key Laboratory of Microscale Magnetic Resonance and School of Physical Sciences, University of Science and Technology of China, Hefei 230026, China}
\affiliation{Anhui Province Key Laboratory of Scientific Instrument Development and Application, University of Science and Technology of China, Hefei 230026, China}
\affiliation{Hefei National Laboratory, University of Science and Technology of China, Hefei 230088, China}
\affiliation{Institute of Quantum Sensing and School of Physics, Zhejiang University, Hangzhou 310027, China}

\begin{abstract}
Quantum friction, \erl{a quantum analog of classical friction,} reduces the performance of quantum machines, \erl{such as heat engines,} and makes them less energy efficient. We here report the experimental realization of \erl{an energy efficient} quantum engine coupled to a quantum battery \el{that stores the produced work}, using a single ion in a linear Paul trap. We first establish the quantum nature of the device by observing nonclassical work oscillations with the number of cycles 
as verified by energy measurements of the battery. We moreover successfully apply shortcut-to-adiabaticity techniques to suppress quantum \erl{friction} \el{and improve work production}. \el{While the average energy cost of the shortcut protocol is only about $3\%$, the work output is enhanced by up to approximately 33$\%$}, making the machine significantly more energy efficient. In addition, we show that the quantum engine consistently outperforms its classical counterpart in this regime. Our results pave the way for energy efficient machines with quantum-enhanced performance.
\end{abstract}


\maketitle
Thermal machines have been the focus of thermodynamics since the very beginning. They play a central role in our society by converting heat into usable energy, such as mechanical work to generate motion \cite{cen01}. In the past decade, heat engines have been successfully miniaturized to the nanoscale \cite{hug02,ros16,lin19,ass19,pet19,hor20,bou20,kla19,kim22,ji22,koc23}, where quantum effects, such as coherent superpositions of states \cite{str17}, are expected to significantly influence their properties at low temperatures. The future development of practical quantum engines faces three critical issues that include  (1) the quest of unambiguous observable signatures of quantum behavior \cite{fri17,wat17,uzd15}, (2) the suppression of detrimental quantum friction mechanisms \cite{kos02,fel03,pla14}, and (3) the determination of regimes of quantum-enhanced performance \cite{scu03,scu11,fra20}.

The beneficial impact of quantum coherence on the characteristics of quantum engines has been reported in some instances \cite{kla19,kim22,ji22}. However, generic experimental means to witness coherent engine operation, \er{irrespective of the working medium employed or the specific type of engine} \er{considered, and without directly affecting the quantum dynamics of the machine}, are missing. In addition, like classical thermal machines \cite{cen01}, quantum heat engines are subjected to dissipative losses \cite{kos02,fel03,pla14}. Quantum friction thus occurs when the Hamiltonian of the working medium of the engine does not commute with that of the external driving \cite{kos02,fel03,pla14}. Short cycle times, associated with fast driving, lead in this case to nonadiabatic transitions that \er{substantially} suppress the work output \cite{kos02,fel03,pla14}.
Lossy quantum devices are not energy efficient, since available resources are not used efficiently. A key task is therefore to design energy efficient machines that deliver more output for similar inputs, without sacrificing power  \cite{ame08}. \er{Energy efficient quantum engines have not been demonstrated experimentally so far.}

We here successfully address all \el{these} three experimental challenges  \el{at the same} time. We realize a two-level quantum engine weakly coupled to a quantum harmonic oscillator battery using a single $^{40}$Ca$^+$  ion in a linear Paul trap \cite{lei03}. The cycle is implemented by driving the ion with \erl{a narrow-linewidth laser controlled by arbitrary waveform generator} \cite{bai13,bow13}, whereas \el{coherent} heating and cooling are achieved using laser \erl{pumping} techniques \cite{lei03}. We demonstrate the quantum nature of the machine by measuring the energy stored in the battery after a variable number of cycles; \el{the engine is not \erl{directly} measured in order to preserve its quantum features.} The energy \el{of the harmonic oscillator battery} does not increase linearly with the cycle number, as for classical engines, but exhibits oscillations that reveal intercycle quantum coherence \cite{wat17}. We further reduce internal quantum dissipation \el{\cite{kos02,fel03,pla14}} by applying a powerful shortcut-to-adiabaticity protocol \cite{den13,cam14,bea16,jar16,cho16,aba17,aba18,cak19,aba19,har20}, known as counterdiabatic driving \cite{cam13,tor13,gue19}. By suppressing nonadiabatic excitations, shortcut-to-adiabaticity methods allow the engineering of adiabatic dynamics in finite time,  \er{as shown in recent experiments \cite{cou08,sch11,bas12,bow12,wal12,zha13,du16,an16,den18,yin22} albeit not for cyclic thermal machines}. \el{Whereas the energetic cost of the counterdiabatic driving is only about $3\%$, we observe an up to \erl{$33.2 \%$} increase of the work output}. Shortcut-to-adiabaticity techniques thus make the engine significantly more energy efficient. Moreover, in this regime, the quantum device \el{produces more work than} its classical counterpart, \er{a clear signature of quantum advantage}.

\begin{figure}[t]
\centering
\begin{tikzpicture}
\node (a) [label={[label distance=.4 cm]152: \textbf{a)}}] at (-0.6,0) {\includegraphics[width=0.329\textwidth]{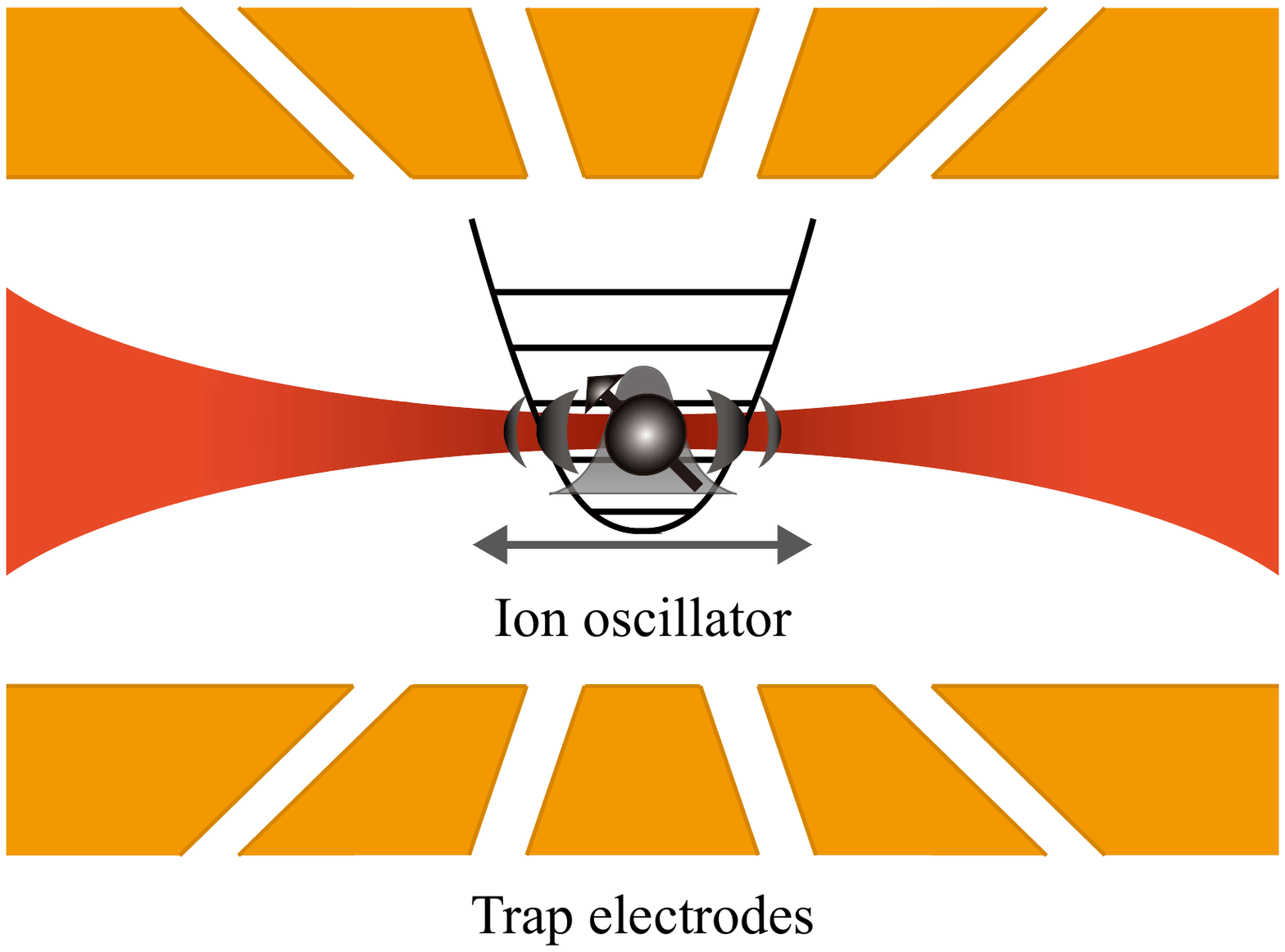}};	\node (a) [label={[label distance=.3 cm]152: \textbf{b)}}] at (-0.37,-5.1)  {\includegraphics[width=0.365\textwidth]{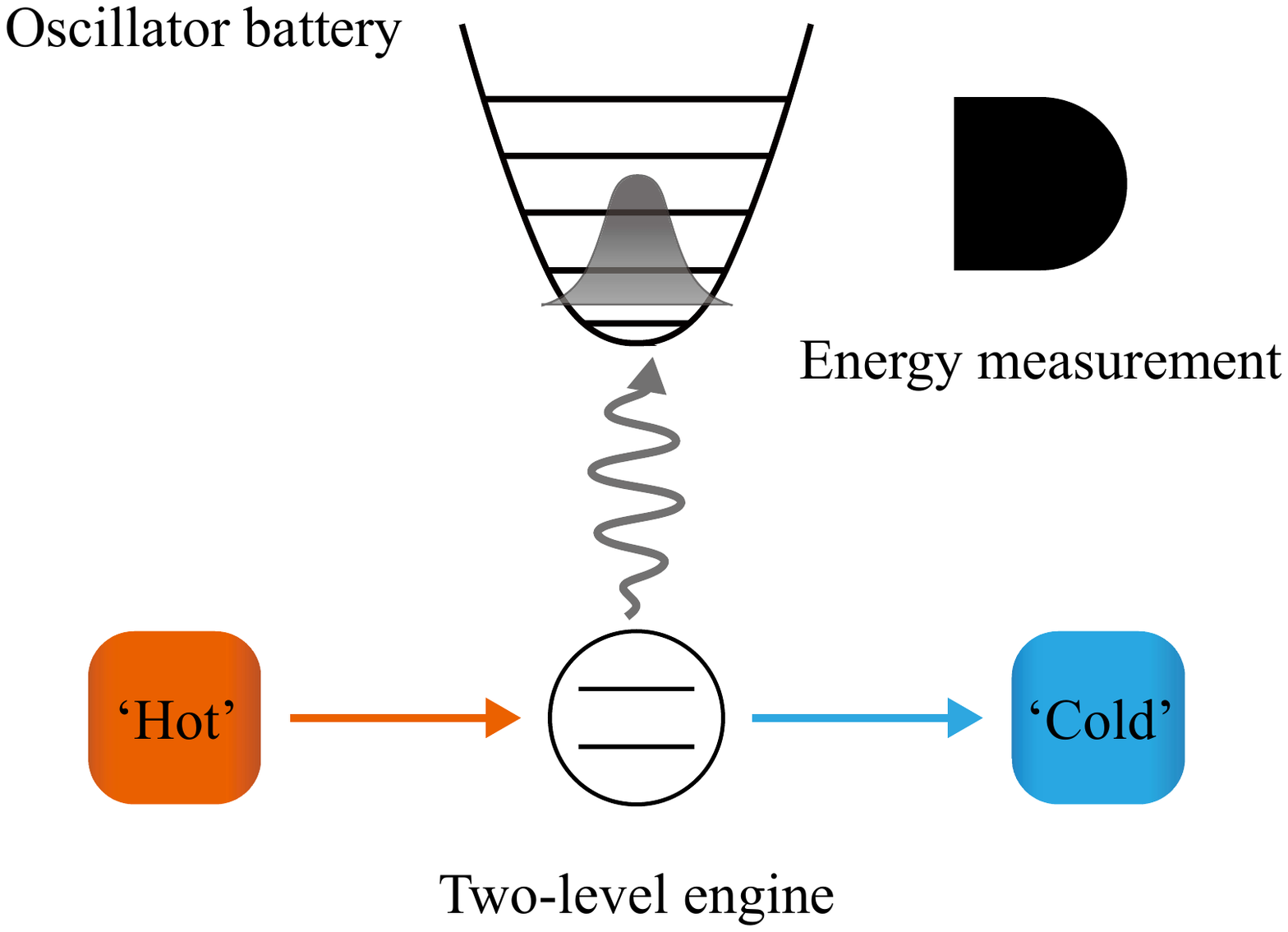}}; 	\node (a) [label={[label distance=.3 cm]152: \textbf{c)}}] at (-0.37,-10.2)  {\includegraphics[width=0.365\textwidth]{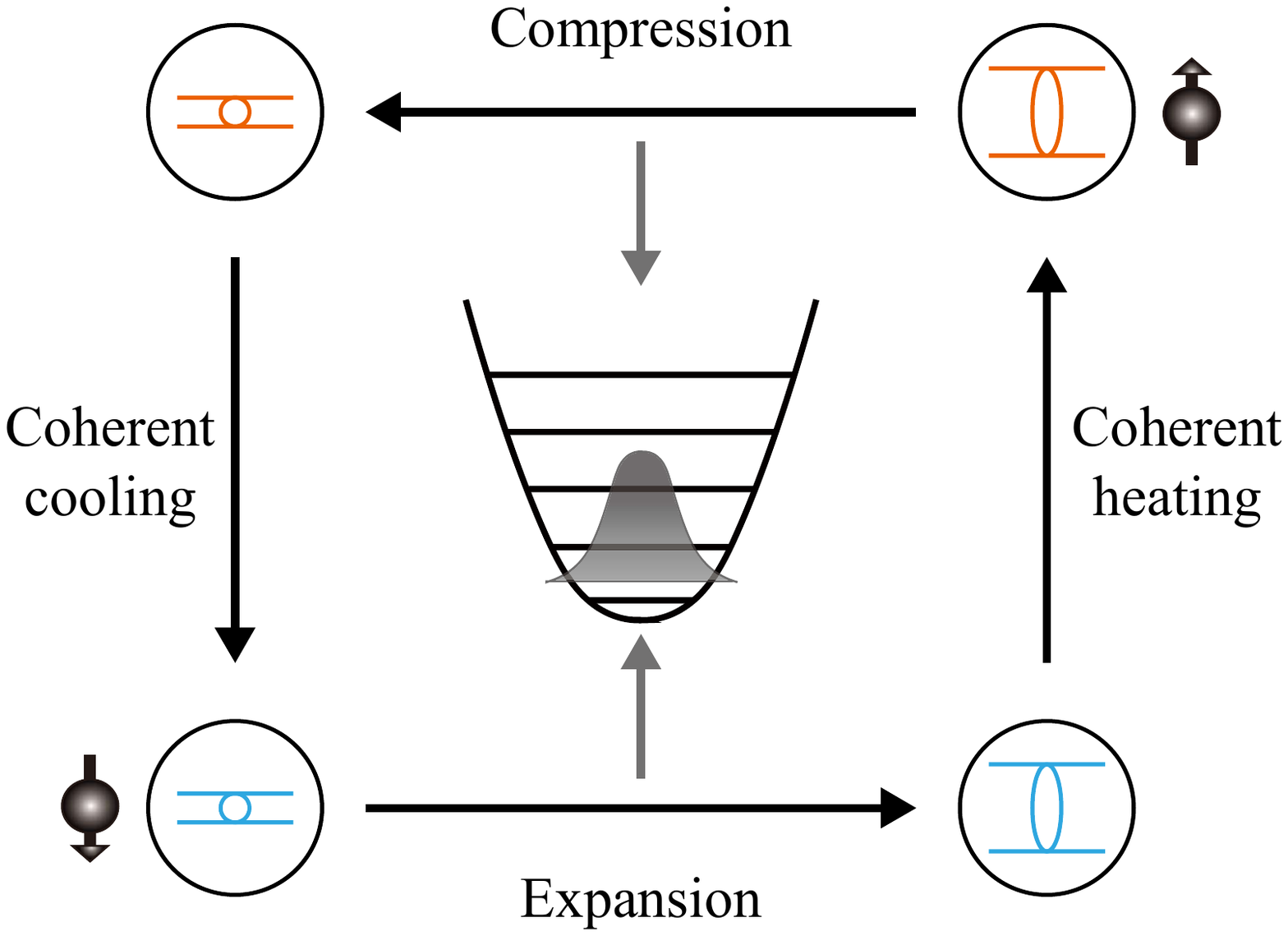}};
\end{tikzpicture}
\caption{\el{Quantum heat engine. a) A single ion trapped in a harmonic potential is subjected to control laser fields to realize the quantum machine. b) A qubit heat engine cyclically operates between cold and hot coherent baths, and stores the produced work in a quantum harmonic oscillator battery, whose energy is measured after a number of cycles. c) The cycle consists of four consecutive steps: isochoric expansion, coherent heating, isochoric compression and coherent cooling.}}
\end{figure}

\emph{Model and experimental setup.}
 We consider a quantum engine ($E$) with a spin-1/2 (with frequency $\Omega$) as its working medium (Fig.~1). The heat engine is coupled to a quantum battery ($B$) that consists of a harmonic oscillator (with frequency $\omega$) that stores the produced work. The corresponding Hamiltonian reads $H={H_E}+H_B+{H_{EB}}$, with ${H_E}=({\Omega}/2)\sigma_x+[{v(t)}/{2}]\sigma_z$ and $H_B=\omega{\textit{a}^\dag}a$ (in units of $\hbar$). The operators $\sigma_{x, y, z}$ denote the Pauli matrices, whereas $a$ and $a^\dagger$ are the usual ladder operators of the harmonic oscillator. \el{The engine Hamiltonian $H_E$ is of the Landau-Zener type, a versatile model of a driven two-level system \cite{iva23}.} The function $v(t)$ is the external driving field. Note that the driving term does not commute with the Hamilton operator of the qubit.  The \er{engine-battery} coupling  is of the form ${H_{EB}}=-({\eta\Omega}/{2})\sin({\omega}t)\sigma_y(a+a^\dag)$, with Lamb-Dicke factor $\eta$ \cite{lei03}.
We \er{concretely} examine a quantum Otto cycle \el{in the diabatic basis of the engine Hamiltonian, that is, in the eigenbasis of $H_E$ with $\omega=0$  \cite{iva23}}, which is composed of the following four steps \cite{kos17}: (1) Unitary expansion during which the spin is driven with the field ${v}(t)=v_0{(2{t}/{\tau})}^2(3-4{t}/{\tau})$ from $t=0$ to ${\tau/2}$, (2) Hot isochore during which the spin is heated into the excited state in a negligible time, (3) Unitary compression with driving field $v(\tau-t)$ from $t=\tau/2$ to ${\tau}$, and (4) Cold isochore, which closes the cycle by cooling the spin back to its ground state in a negligible time (Fig.~1). \el{These four branches correspond to the cycle of a quantum engine coupled to coherent reservoirs \cite{scu03,scu11,guf19,aim23} in the adiabatic basis of the Landau-Zener Hamiltonian, that is, in the instantaneous eigenbasis of $H_E$  \cite{iva23}}.

\begin{figure}[t!]
	\begin{center}
	\includegraphics[width=\columnwidth]{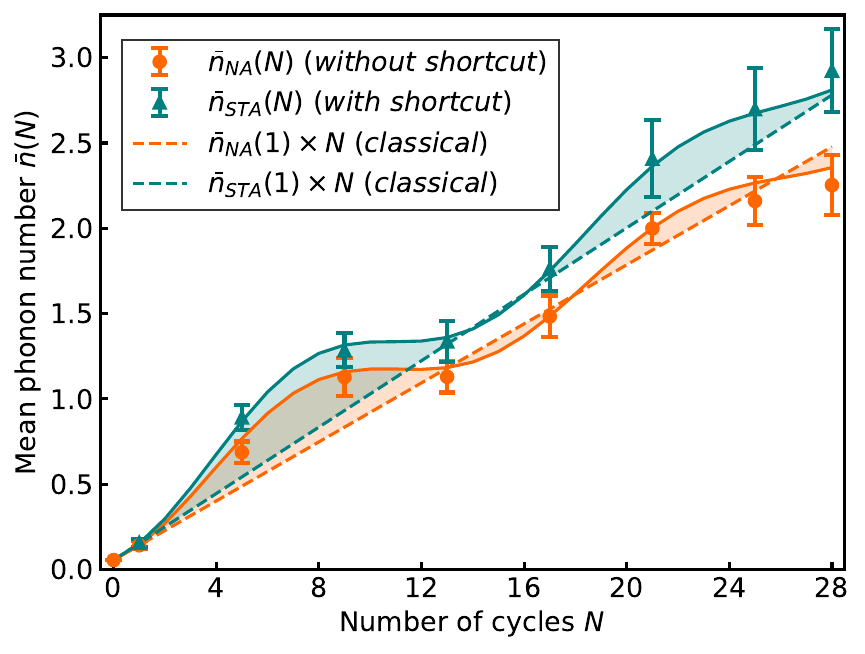}
	\caption{Quantum signature in the engine work output. The mean phonon number $\Bar{n}_{NA}(N)$ determined by an energy measurement of the quantum battery after $N$ cycles exhibits quantum oscillations (orange dots). For a classical engine, the work output scales linearly with $N$ (orange dashed line). With counterdiabatic driving, quantum friction is suppressed and the work output $\Bar{n}_{STA}(N)$ is increased (green triangles) above the classical limit (green dashed line). In both cases, good agreement with numerical simulations (solid lines) is obtained. Parameters are $v_0=\omega=2\pi\times0.075\ \rm{MHz}$ and $\tau=119\ \rm{\mu{s}}$. Error bars correspond to one standard deviation.
}
	\label{fig:figure2_scanN}
	\end{center}
\end{figure} 

In our experiment, we store a single $^{40}\rm{Ca}^+$ ion in a linear Paul trap with an ambient magnetic field of $0.538\ \rm{mT}$, with a motional frequency of $\omega_m =2\pi\times 2.0~\rm{MHz}$ for the axis of interest. The states of the quantum engine are defined by the $D_{5/2}$ ($S_{1/2}$) as the spin-up $\ket{\uparrow}$ (spin-down $\ket{\downarrow}$) state, respectively. We drive between spin states with a near resonant quadruple transitions by a laser beam of approximately 729~nm, corresponding to a Rabi frequency of $\Omega=2\pi\times0.159\ \rm{MHz}$. We apply two additional laser beams detuned from the quadruple resonance by approximately $\pm\omega_m$, thus forming the coupling between the engine and battery. With proper choice of laser frequencies, a desired Hamiltonian is formed after rotating wave approximation, where
the effective harmonic oscillator (with eigenstates $\{\ket{n}\}$) has a frequency \el{of} $\omega=2\pi\times0.075\ \rm{MHz}$, and $v_0=2\pi\times0.075\ \rm{MHz}$. In the experiment, we prepare the system by first cooling the ion to the Doppler limit by driving the $S_{1/2}$ to $P_{1/2}$ dipole transition. We then apply resolved-sideband cooling on the $S_{1/2}$ to $D_{5/2}$ quadrupole transition to cool the ion motion to the ground state \el{$\ket{n=0}$} and initialize the spin to its ground state $\ket{\downarrow}$ \cite{Phys.Rev.Lett.83.4713}. Expansion and compression strokes of the quantum Otto cycle \el{in the diabatic basis} are implemented with the help of an arbitrary wave generator \cite{bai13,bow13} that employs a three-color light field to generate the driving function $v(t)$ with arbitrary amplitude and frequency patterns. Heating is realized by optically pumping the ion into the excited state $\ket{\uparrow}$. The work produced during each closed cycle of the engine is stored in the harmonic oscillator, increasing the mean phonon number $\bar n$. At the end of the strokes with desired number of cycles, we perform laser thermometry of the battery by mapping the information to the spin, followed by fluorescence detection. Details of the above processes are available in the Supplementary Information.

\begin{figure}[t!]
	\begin{center}
	\includegraphics[width=1\columnwidth]{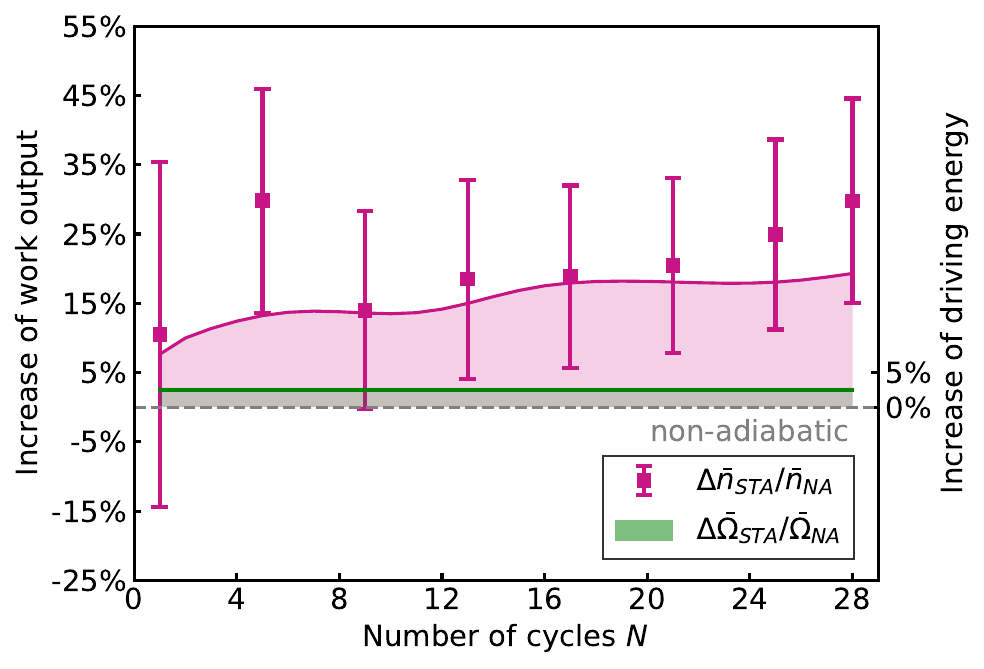}
	\caption{Energy efficient quantum machine. The relative increase of work output, $\Delta \Bar n_{STA}/ \Bar{n}_{NA}$, in the presence of counterdiabatic driving is between 10.5(24.9)$\%$ and 29.8(14.8)$\%$ (pink squares), depending on the cycle number $N$. By contrast, the average energetic cost of the shortcut protocol, $\Delta \Bar \Omega_{STA}/ \Bar{\Omega}_{NA}$, is only about 2.6(0.2)$\%$ (green bar). Available resources are therefore more efficiently used. Solid lines show numerical simulations. Error bars correspond to one \mbox{standard deviation}.}
	\label{fig:figure3_CDcost}
	\end{center}
\end{figure}

\emph{Signature of intercycle quantum coherence.}
We begin by investigating the quantum features of the device by performing stroboscopic  measurements of the energy deposited into the quantum battery \cite{wat17}. To that end, we  determine the (nonadiabatic) mean phonon number $\bar n_{NA}(N)$ of the harmonic oscillator after a variable number $N$ of cycles (up to $N=28$) using standard ion trap techniques \cite{lei03} (Supplementary Information). We observe a periodic oscillatory behavior as a function of $N$ (orange dots in Fig.~2) that clearly reveals intercycle quantum coherence. For a classical heat engine, the work output indeed scales linearly with \el{the cycle number} $N$ (orange dashed line), \el{since no intercycle coherence is built up} \cite{wat17}. In the quantum regime, the performance of the engine strongly depends on whether the energy of the battery  is measured after every cycle (which destroys its intercycle  coherence) or \el{only after $N$ cycles (which preserves intercycle coherence)}, in stark contrast to  a classical machine. We obtain excellent agreement between measured data (orange dots) and numerical simulations (orange line) (Supplementary Information). We emphasize that the oscillations of the work output seen in Fig.~2 cannot be reproduced through a purely \mbox{classical driving \cite{wat17}}.

\begin{figure}[t!]
	\begin{center}
	\includegraphics[width=1\columnwidth]{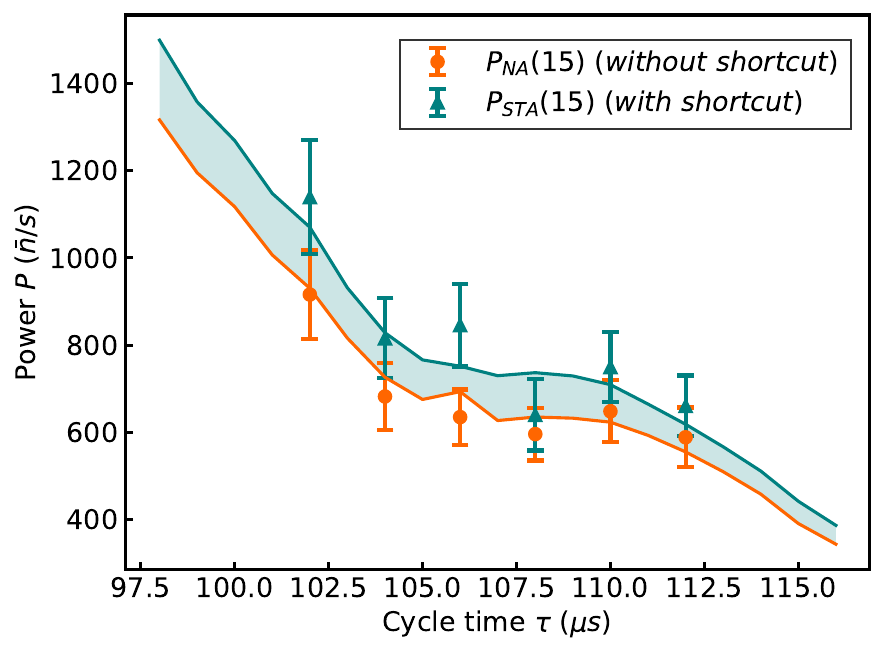}
	\caption{Shortcut enhanced power output. Power output of the quantum engine, $P=\bar n/(N\tau)$, for various cycle times and fixed cycle number $N=15$ without counterdiabatic driving (orange dots) and with counterdiabatic driving (green triangles). The power output increases when quantum dissipation is suppressed by the shortcut-to-adiabaticity protocol. Good agreement is obtained with numerical simulations (solid lines). Error bars correspond to one standard deviation.}
	\label{fig:figure4_power}
	\end{center}
\end{figure} 

\emph{Counterdiabatic suppression of quantum friction.} Another consequence of  quantum coherence is quantum friction which occurs when the Hamiltonian of the working medium does not commute with the external driving Hamiltonian \cite{kos02,fel03,pla14}, as is the case in our experiment. We next use a shortcut-to-adiabaticity scheme \cite{tor13,gue19} to reduce such quantum dissipation and improve the performance of the quantum machine. \el{As a matter of simplicity and practicality, we do not aim at determining the optimal shortcut protocol, which would require precise knowledge of the open nonunitary  dynamics of the spin engine coupled to the external battery. \er{The needed continuous monitoring of the time evolution} would perturb the operation of the quantum machine.} We \el{instead} concretely add a counterdiabatic Hamiltonian   \cite{cak19,ber09,tak13}
  \begin{equation}
H_{CD}(t)=-\frac{1}{2}\frac{\Omega\times \dot v(t)}{{\Omega}^2+{v(t)}^2}\sigma_y = \frac{\Omega_{CD}(t)}{2} \sigma_y,
\label{Eq:Ham_2}
\end{equation}
to the engine Hamiltonian $H_E$ in order to suppress detrimental nonadiabatic transitions \cite{tor13,gue19}, \el{and reduce coherent oscillations along  the $\sigma_y$ direction (Supplementary Information)}. We experimentally realize Eq.~\eqref{Eq:Ham_2} by adding  another \el{spin driving} component to the three-color light field with a $-\pi/2$ phase difference with respect to ${H_E}(0) $ to obtain an effective $\sigma_y$ operator that is controlled by the {arbitrary wave generator}.

\begin{figure}[t!]
	\begin{center}
	\includegraphics[width=1\columnwidth]{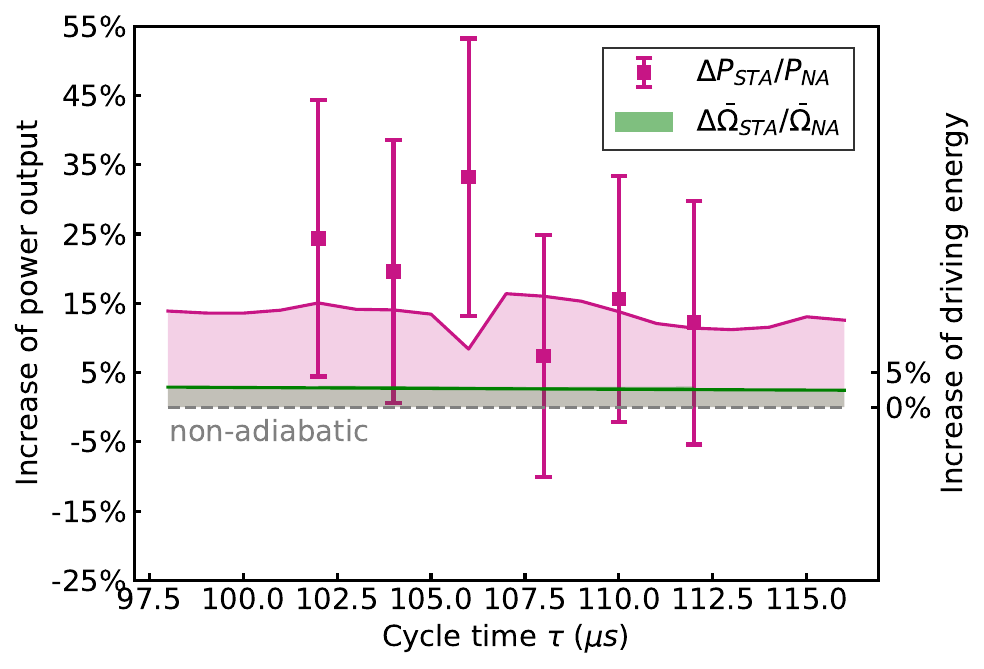}
	\caption{Energy efficient quantum  machine. The relative increase of power output, $\Delta P_{STA}/ P_{NA}$, in the presence of counterdiabatic driving is  between 7.4(17.5)$\%$ and 33.2(20.0)$\%$  (pink squares), depending on the cycle time $\tau$. By contrast, the average energetic cost of the shortcut protocol, $\Delta \Bar \Omega_{STA}/ \Bar{\Omega}_{NA}$,  varies between 2.7(0.2)$\%$ and 2.9(0.2)$\%$ (green bar). Available resources are thus more efficiently used. Solid lines show  numerical simulations. Error bars correspond to one standard deviation.}
	\label{fig:figure5}
	\end{center}
\end{figure}

Figure 2 shows the results of the stroboscopic energy measurement of the  quantum battery in the presence of the  counterdiabatic driving \eqref{Eq:Ham_2} (green triangles). A robust enhancement of the work output $\bar n_{STA}(N)$ is seen, which steadily increases with the number of cycles, up to 29.8(14.8)$\%$ for $N=28$. The counterdiabatic driving hence plays the role of a quantum lubricant \cite{fel06}. We again obtain excellent agreement between  data (green triangles) and numerical simulations (green line). Remarkably, all  data points lie above the classical limit  \el{(with no intercycle coherence)} which corresponds to a linear scaling with the cycle number (green dashed line). We can therefore conclude  that the quantum machine here outperforms its classical counterpart. This represents a distinctive sign of quantum advantage.

We additionally evaluate the average energetic cost of the shortcut-to-adiabaticity protocol during  expansion and compression steps by measuring the Rabi frequency $\Omega_{CD}(t)$ of the counterdiabatic driving \eqref{Eq:Ham_2} and calculating the time average, $\Bar{\Omega}_{CD}=\frac{1}{\tau}\int_{0}^{\tau} \Omega_{CD}(t) dt$, over one cycle (Supplementary Information). The latter quantity is proportional to the intensity of the laser needed to implement the shortcut Hamiltonian. Compared with the mean Rabi frequency $\Bar{\Omega}_{NA}$ of the carrier laser, which is always required to drive the quantum spin engine, we find a relative increase  of $\Delta \Bar \Omega_{STA}/ \Bar{\Omega}_{NA} = (\Bar \Omega_{STA} - \Bar{\Omega}_{NA})/ \Bar{\Omega}_{NA} $ of only 2.6(0.2)$\%$ (green bar in Fig.~3). This value should be contrasted with the relative increase of work output achieved by the shortcut-to-adiabaticity protocol, $\Delta \Bar n_{STA}/ \Bar{n}_{NA} = (\Bar n_{STA} - \Bar{n}_{NA})/ \Bar{n}_{NA}$, which lies between 10.5(24.9)$\%$ and 29.8(14.8)$\%$,  with a mean of 20.8(5.6)$\%$, (pink squares in Fig.~3). The quantum lubrification brought about by the counterdiabatic driving  hence leads to a significant enhancement of the performance of the engine (of almost one order of magnitude  compared to  the relatively small energetic cost of the shortcut). This makes the superadiabatic quantum  engine more energy efficient, since available resources are exploited much  more efficiently after the decrease of quantum friction mechanisms.

Figure 4 displays an analysis of the impact of the counterdiabatic driving  \eqref{Eq:Ham_2} on the power output, $P=\bar n/(N\tau)$, of the quantum  engine by varying the cycle time $\tau$ for a constant cycle number $N=15$ (the background heating rate 
from the trap has been taken into account and subtracted). We note that the nonadiabatic power output $P_{NA}$, without shortcut, increases when the cycle time decreases, as expected, and that  the power output $P_{STA}$,  in the presence of the shortcut-to-adiabaticity driving,  is augmented by an average  value of 18.7(7.6)$\%$, from 12.2(17.6)$\%$ for $\tau = 112\mu$s to 24.4(20.0)$\%$ for $\tau = 102\mu$s. The associated relative power enhancement, $\Delta P_{STA}/ P_{NA} = (P_{STA} - P_{NA})/ P_{NA} $ (pink squares), as well as the corresponding average energetic cost, $\Delta \Bar \Omega_{STA}/ \Bar{\Omega}_{NA}$ (green bar),
of the shortcut-to-adiabaticity protocol are presented as a function of $\tau$ in Fig.~5. The latter quantity increases when the cycle duration is reduced, from 2.7(0.2)$\%$ for $\tau = 112\mu$s to 2.9(0.2)$\%$ for $\tau = 102\mu$s, due to the fact that more nonadiabatic transitions have to be compensated; as before, solid lines show the numerical simulations of the motor. Remarkably, the associated relative power enhancement $\Delta P_{STA}/ P_{NA}$ is obtained up to 33.2(20.0)\%, in stark contrast to the average energetic cost $\Delta \Bar \Omega_{STA}/ \Bar{\Omega}_{NA}$ of up to 2.9(0.2)\%. For all values of the cycle time $\tau$, the mean increase of the power output hence largely exceeds the average energetic cost of the counterdiabatic driving, underscoring the improved energy efficiency of the superadiabatic quantum heat engine.

\emph{Conclusions.} Assessing the quantum properties of quantum machines is a challenging task, since direct measurements will usually disrupt their delicate quantum features. We have experimentally studied the characteristics of a single ion quantum heat engine realized in a linear Paul trap by measuring the energy stored in a quantum harmonic oscillator battery. We have observed nonclassical oscillations of the work output with the cycle number, thus revealing intercycle quantum coherence. We have furthermore successfully suppressed quantum friction with the help of shortcut-to-adiabaticity methods that act as a quantum lubricant. In doing so, we have not only increased its performance by about 20$\%$ on average, but also significantly increased its energy efficiency, since the net gain is much larger than the thermodynamic cost of the shortcut protocol. We have moreover established its quantum advantage by showing that it consistently outperforms its classical counterpart. Our findings indicate that quantum-enhanced performance can be advantageously combined with energy efficiency. In view of the versatility of our results, we expect them to be of importance for the design of  \er{energy efficient quantum thermal machines, such as heat engines, refrigerators and pumps \cite{kos14}, of energy efficient quantum information engines \cite{fad23}, as well as of energy efficient quantum technologies \cite{auf22}.} 

\emph{Acknowledgments.} The USTC team acknowledges support from the National Natural Science Foundation of China (Grant No. 92165206, 11974330),  Innovation Program for Quantum Science and Technology (Grant No. 2021ZD0301603), the USTC start-up funding, and the Fundamental Research Funds for the Central Universities, and Hefei Comprehensive National Science Center. K.R. further acknowledges support from CAS-PIFI (2021PM0049) and China Ministry of Education Funding for Cultivating Key Projects in the Important Directions of Basic Scientific Research (WK3540000004). E.L. is supported by the German Science Foundation (DFG) (Grant No. FOR 2724).\\




\newpage
\appendix
\setcounter{figure}{0}
\setcounter{equation}{0}

\renewcommand*{\thefigure}{S\arabic{figure}}
\renewcommand*{\theequation}{S\arabic{equation}}

\maketitle
{\centering\section{Supplementary Information: An energy efficient quantum-enhanced machine}}

\section{Theoretical Details}

\subsection{Derivation of the Hamiltonian}
We experimentally implement the Hamiltonian of a spin quantum heat engine system (in units of $\hbar$):
\begin{equation}
H(t)={H_E}(t)+H_B+{H_{EB}}(t),
\label{Eq:Ham_1}
\end{equation}
with
\begin{gather*}
{H_E}(t)=\frac{\Omega}{2}\sigma_x+\frac{v(t)}{2}\sigma_z,\\
H_B=\omega{\textit{a}^\dag}a,\\
{H_{EB}}(t)=-\frac{\eta\Omega}{2}\sin({\omega}t)\sigma_y(a+{\textit{a}^\dag})
\end{gather*}
{\noindent}which describes an ensemble consisting of a spin-$\frac{1}{2}$ particle and its momentum phonon system. This Hamiltonian can be constructed by the eigen-Hamiltonian of an ion trapped in harmonic confinement with the ion-laser interaction Hamiltonian under rotating-wave approximation (RWA) \cite{lei03}. The energy of a two-level ion trapped in a harmonic confinement is described by the Hamiltonian
\begin{equation}
H_0=\frac{\omega_0}{2}\sigma_z+\omega_z({\textit{a}^\dag}a+\frac{1}{2})
\label{Eq:Ham_2}
\end{equation}
{\noindent}where $\omega_z$ is the ion’s motional frequency along the $z$-direction of the potential well, and $\omega_0$ is the unperturbed two-level system (TLS) transition frequency. A laser light field with frequency $\omega_L$ and phase $\phi_L$ induces a perturbation described by the Hamiltonian
\begin{equation}
H_L=\frac{\Omega}{2}\sigma_x(e^{i\eta(a+{\textit{a}^\dag})}e^{-i({\omega_L}t+\phi_L)}+h.c.)
\label{Eq:Ham_3}
\end{equation}
{\noindent}where the Rabi frequency $\Omega$ represents the coupling strength between the laser field and the bare two-level atomic transition. An additional term $\exp(ik\hat{x})=\exp\{i\eta(a+{\textit{a}^\dag})\}$, where the Lamb-Dicke parameter $\eta=k{x}_{0}\cos\theta$, describes the interaction between laser and a trapped ion with additional oscillating motion. With the overall Hamiltonian $H=H_0+H_L$, a transformation with $U^{spin}=\exp(-i{H}_{0}^{spin}t)$, where ${H}_{0}^{spin}=\frac{\omega_0-v}{2}\sigma_z$, into the interaction picture yields:
\begin{equation}
    \begin{split}
        H_{int}^{spin}&=\frac{v(t)}{2}\sigma_z+\omega_z({\textit{a}^\dag}a+\frac{1}{2})\\
        &+\frac{\Omega}{2}(e^{i\eta(a+{\textit{a}^\dag})}e^{-i({\delta}t+\phi_L)}\sigma_{+}+h.c.)
    \end{split}
\label{Eq:Ham_4}
\end{equation}
{\noindent}where $\delta=\omega_L-(\omega_0-v)$ denotes the detuning of the laser frequency from the atomic transition. Here, a RWA has been performed and all terms rotating at the sum frequency $\omega_L+(\omega_0-v)$ are neglected, as they average out over the time scale of $\delta$. Then another transformation with $U^{motion}=\exp(-i{H}_{0}^{motion}t)$, where ${H}_{0}^{motion}=(\omega_z-\omega){\textit{a}^\dag}a=\omega_{z}^{'}{\textit{a}^\dag}a$, is performed to further simplify the interaction Hamiltonian:
\begin{equation}
    \begin{split}
        H_{int}&=\frac{v(t)}{2}\sigma_z+\omega{\textit{a}^\dag}a\\
        &+\frac{\Omega}{2}(e^{-i({\delta}t+\phi_L)}\sigma_{+}\exp\{i{\eta}(a{e}^{-i\omega_{z}^{'}t}+{\textit{a}^\dag}{e}^{i\omega_{z}^{'}t})\}+h.c.)
    \end{split}
\label{Eq:Ham_5}
\end{equation}
{\noindent}In the Lamb-Dicke regime ${\eta}^{2}(2\bar{n}+1)\ll1$, where $\bar{n}$ is the average phonon number of the harmonic oscillator, the ion’s motional wave packet is confined to an extent which is much smaller than the laser’s wavelength. This Lamb-Dicke approximation can further simplify Eq.~\eqref{Eq:Ham_5} by Taylor expanding it to:
\begin{equation}
    \begin{split}
        H_{int}&=\frac{v(t)}{2}\sigma_z+\omega{\textit{a}^\dag}a\\
        &+\frac{\Omega}{2}(e^{-i({\delta}t+\phi_L)}\sigma_{+}\{1+i{\eta}(a{e}^{-i\omega_{z}^{'}t}+{\textit{a}^\dag}{e}^{i\omega_{z}^{'}t})\}+h.c.)
    \end{split}
\label{Eq:Ham_6}
\end{equation}

\begin{figure}[t!]
	\begin{center}
	\includegraphics[width=1\columnwidth, angle=90]{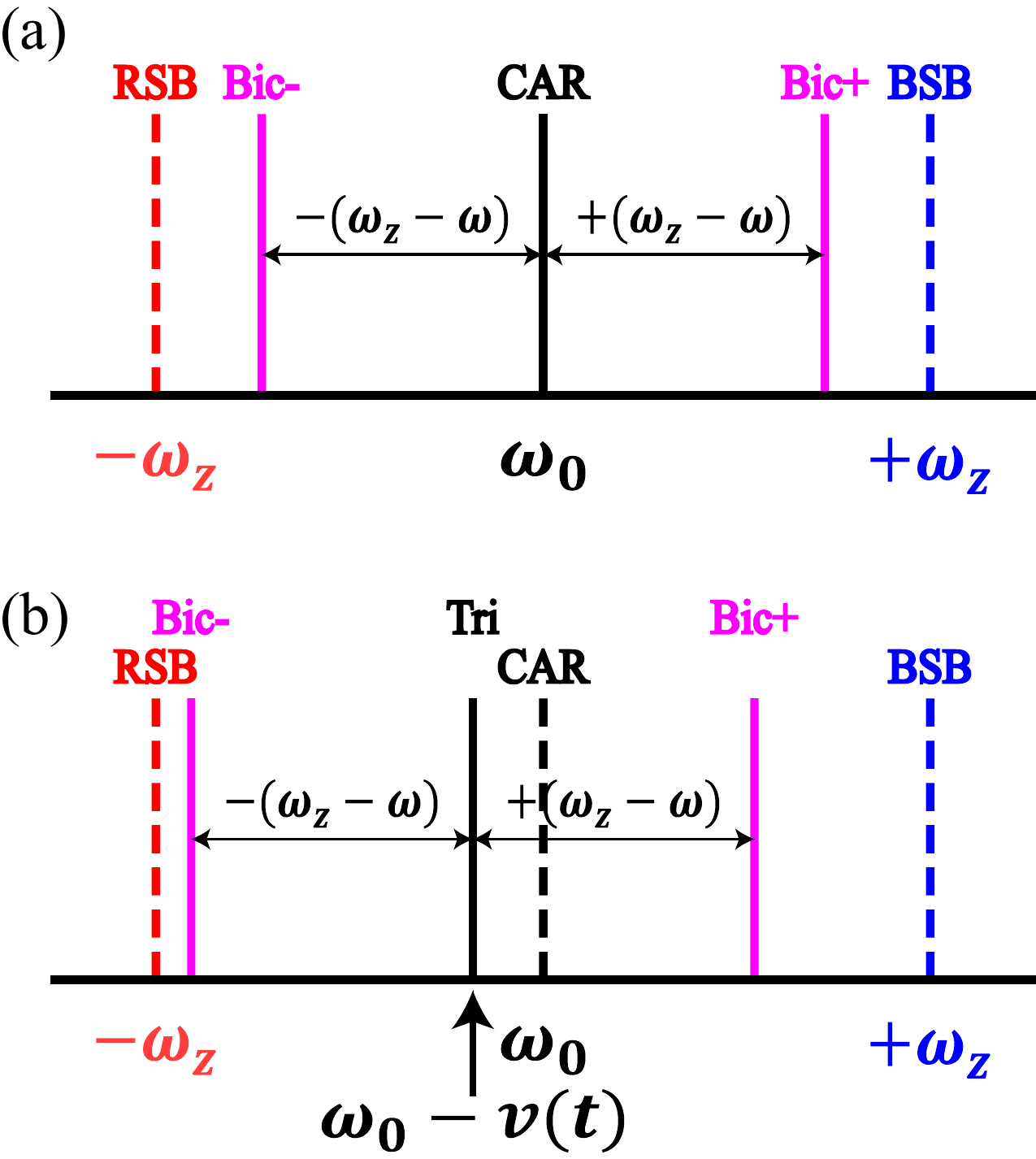}
	\caption{Illustration of the three-color light field used in the experiment. The dashed lines show the unperturbed frequency of the TLS transition and harmonic oscillator, while the solid lines show the frequency of the three-color light field. (a) The three-color light field at the start ($t=0$) and end ($t=\tau$) of one Otto cycle. The bichromatic light field, denoted as Bic+ and Bic-, has a symmetric detuning $\omega$ from the harmonic oscillator mode $\omega_z$, leading to the transformation in Eq.~\eqref{Eq:Ham_5}. (b) The three-color light field at time $t$ during expansion and compression strokes. An additional detuning $-v(t)$ is induced onto the center frequency of the three-color light field, known as a center-line detuning, leading to the transformation in Eq.~\eqref{Eq:Ham_4}.}
	\label{fig:figureS1_laser}
	\end{center}
\end{figure}

In the experiment, we apply a three-color light field in the same direction to an ion to induce the total Hamiltonian $H_{L}^{total}=H_{L}^{carrier}+H_{L}^{red}+H_{L}^{blue}$, as shown in Fig.~\ref{fig:figureS1_laser}. The frequency and phase relationship between them is $\delta^{carrier}=0$, $\delta^{blue}=-\delta^{red}=\omega_{z}^{'}$ and $\phi_{L}^{carrier}=\phi_{L}^{blue}=\phi_{L}^{red}=0$. An intensity modulation with frequency $\omega$ is performed onto the red-blue sideband bicolor light $\Omega^{blue}=\Omega^{red}={\Omega}\sin({\omega}t)$, while the amplitude is the same with carrier light $\Omega^{carrier}=\Omega$. The main reason for using this intensity modulation function is to ensure that the average phonon number of the harmonic oscillator can also be increased accordingly under the new rotating frame, while the coherence of the spin quantum heat engine itself is unaffected \cite{PhysRevLett.118.050601,Phys.Rev.E.78.011115}, so that the work produced by the engine can be distinguished from the noise of background heating. After taking into account all the three frequency components, a RWA is performed, and all terms rotating at $\omega_{z}^{'}$ and its higher orders are neglected. The total interaction Hamiltonian can be finally modified to:
\begin{equation}
    \begin{split}
        H_{int}^{total}&=\frac{v(t)}{2}\sigma_z+\omega{\textit{a}^\dag}a+\frac{\Omega^{carrier}}{2}\sigma_x\\
        &+\frac{\Omega^{blue}}{2}(i\eta{\sigma}_{+}{\textit{a}^\dag}+h.c.)+\frac{\Omega^{red}}{2}(i\eta{\sigma}_{+}a+h.c.)\\
        &=\frac{v(t)}{2}\sigma_z+\omega{\textit{a}^\dag}a+\frac{\Omega}{2}\sigma_x-\frac{\eta\Omega}{2}\sin({\omega}t)\sigma_y(a+{\textit{a}^\dag})
    \end{split}
\label{Eq:Ham_7}
\end{equation}
{\noindent}It can be seen that Eq.~\eqref{Eq:Ham_7} is consistent with Eq.~\eqref{Eq:Ham_1}.

\subsection{Shortcuts to adiabaticity for two-level system}
For an arbitrary time-dependent isolated Hamiltonian $H_{0}(t)$ with instantaneous eigenstates $\{\ket{j(t)}\}$ and energies $\{E_{j}(t)\}$, from Berry’s formulation \cite{J.Phys.A.42.365303}, the resulting counterdiabatic Hamiltonian $H_{CD}(t)$ is:
\begin{equation}
H_{CD}(t)=i\sum_{j}(\ket{\partial_{t}j}\bra{j}-\braket{j|\partial_{t}j}\ket{j}\bra{j})
\label{Eq:Ham_8}
\end{equation}
{\noindent}Note that the eigen-Hamiltonian of the spin quantum engine is ${H_E}(t)=\frac{\Omega}{2}\sigma_x+\frac{v(t)}{2}\sigma_z$, and it corresponds to the Landau-Zener  model which is
well studied in many physical settings \cite{IVAKHNENKO20231}. The counterdiabatic Hamiltonian of the LZ model can be directly written as \cite{J.Phys.A.42.365303,Phys.Rev.E.99.032108}:
\begin{equation}
\begin{split}
    H_{CD}(t)&=\frac{1}{2}\frac{\Dot{\Omega}{v(t)}-\Omega\Dot{v(t)}}{{\Omega}^2+{v(t)}^2}\sigma_y\\
    &=-\frac{1}{2}\frac{\Omega\times\Dot{v(t)}}{{\Omega}^2+{v(t)}^2}\sigma_y
\end{split}
\label{Eq:Ham_9}
\end{equation}
{\noindent}To ensure that the effective Hamiltonian $H_{STA}(t)=H(t)+H_{CD}(t)$ equals the original Hamiltonian $H(t)$ [Eq.~\eqref{Eq:Ham_1}] at the start and end of the protocol, we impose a condition on the form of $v(t)$: $\Dot{v(0)}=\Dot{v(\tau/2)}=0$. Here, for simplicity, we select the following time-dependence profile of ${v}(t)=v_0{(\frac{t}{\tau/2})}^2(3-2\frac{t}{\tau/2})$ to satisfy this boundary condition, while the energy gap of the TLS at the start and end of the protocol is still $E(0)=\Omega$ and $E(\tau/2)=\sqrt{\Omega^2+v_0^2}$. In the experiment, $H_{CD}$ [Eq.~\eqref{Eq:Ham_9}] can be constructed by adding a laser field with $\delta^{CD}=\delta^{carrier}=0$, $\phi_{L}^{CD}=-\pi/2$ and $\Omega^{CD}=\frac{\Omega\times\Dot{v(t)}}{{\Omega}^2+{v(t)}^2}$ [Eq.~\eqref{Eq:Ham_6}].

\section{Experimental Details}

\subsection{Experimental setup and data processing}
\begin{figure}[t!]
	\begin{center}
	\includegraphics[width=1\columnwidth]{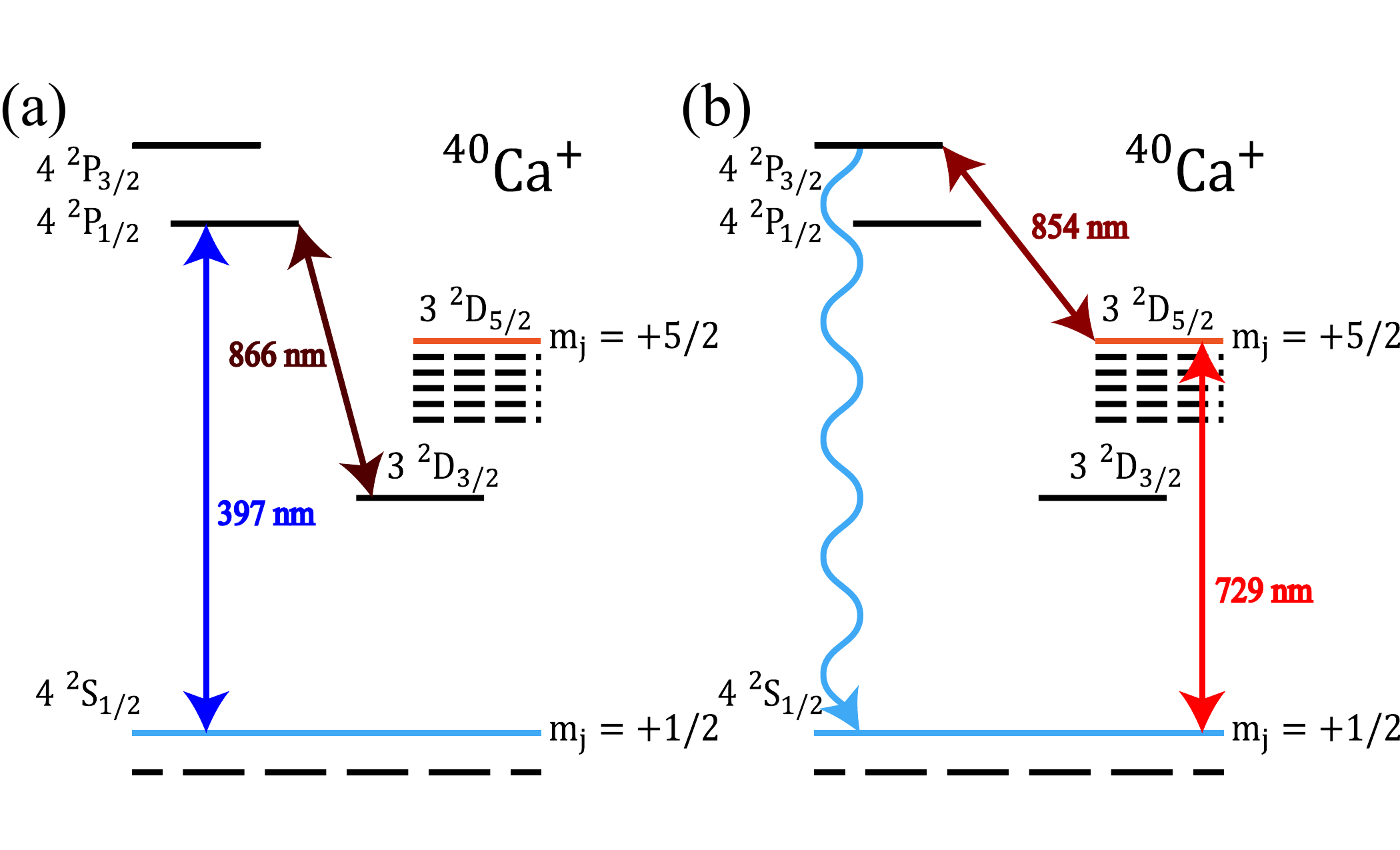}
	\caption{Illustration of the $^{40}\rm{Ca}^+$ ion energy levels and transitions. (a) The population of the ground state $S_{1/2}$ is detected via electron shelving. (b) The $854\ \rm{nm}$ laser is used to quench the metastable state $D_{5/2, m_j=+5/2}$, thus reset the spin state to ground state $S_{1/2, m_j=+1/2}$. The $729\ \rm{nm}$ laser is used for initial state preparation and engine driving.}
	\label{fig:figureS2_ca_level}
	\end{center}
\end{figure}

To simulate such a quantum heat engine system, we trap a single $^{40}\rm{Ca}^+$ ion in a linear Paul trap with the ambient magnetic field of $0.538\ \rm{mT}$. The harmonic oscillator eigenstates of $^{40}\rm{Ca}^+$ ion along the $z$-axis of the potential well, denoted as $\{\ket{n}\}$, are utilized as the eigenstates of the phonon system. Its oscillation frequency within the potential well is denoted as $\omega_z$. For $^{40}\rm{Ca}^+$ ion, the population of ground state $4s~^2 S_{1/2, m_j=+1/2}$, which is selected as the spin-down state $\ket{\downarrow}$, can be detected by means of electron shelving \cite{Phys.Rev.Lett.83.4713}. Here, $397\ \rm{nm}$ laser light excites the transition between $S_{1/2}-P_{1/2}$. If the valence electron is in the $S_{1/2}$ state, photons are scattered and collected by a photonmultiplier tube (PMT), whereas if the electron is in the $D_{5/2}$ state, which is selected as the spin-up state $\ket{\uparrow}$, no photons are scattered. A laser at $866\ \rm{nm}$ serves to repump the ion via $P_{1/2}-D_{3/2}$ during electron shelving, thus closing the fluorescence excitation cycle, as shown in Fig.~\ref{fig:figureS2_ca_level}(a). For each data point of spin state, the preparation, evolution, and detection are repeated 200 times. We simulate the error of spin state detection by assuming the normal distribution of the raw data, and sample 200 times from the raw data, leading to an estimation of the standard deviation of the spin state population.

During the heating and cooling strokes of one Otto cycle, the state of the spin needs to be reset and prepared in a negligible time to thermalize with the heat baths. In the experiment, the $854\ \rm{nm}$ laser is used to excite the transition between $D_{5/2}-P_{3/2}$, then the spontaneous emission of $P_{3/2}$ state causes the valence electron to return to $S_{1/2}$ state, thus resetting the spin state to $\ket{\downarrow}$ state while leaving the phonon state unaffected. Since we select the metastable state $3d~^2 D_{5/2, m_j=+5/2}$ and ground state $4s~^2 S_{1/2, m_j=+1/2}$ as the spin-up ($\ket{\uparrow}$) and spin-down ($\ket{\downarrow}$) states respectively, the selection rules ensure that the valence electron in the $D_{5/2, m_j=+5/2}$ state will only be excited to $P_{3/2, m_j=+3/2}$ state, and then de-excite to $S_{1/2, m_j=+1/2}$ state, thus ensuring that no spin population leakage to other sub-levels during the multiple cycles, as shown in Fig.~\ref{fig:figureS2_ca_level}(b). After resetting the spin state with a $854\ \rm{nm}$ laser, when it is needed to prepare the initial spin state according to the temperature of the heat baths, the $729\ \rm{nm}$ laser pulses of different duration times can be used to adjust the population of $\ket{\uparrow}$ state. In the experiment, for simplicity, we always set the hot equilibrium state to the $\ket{\uparrow}$ state, and the cold equilibrium state to the $\ket{\downarrow}$ state. The resetting and initial state preparation of the spin can be finished in less than $5\ \rm{{\mu}s}$, thus ensuring that the time spent in the heating and cooling strokes is negligibly small compared to the total time of one Otto cycle, which is usually around $100\ \rm{{\mu}s}$.

\begin{figure}[t!]
	\begin{center}
	\includegraphics[width=1\columnwidth]{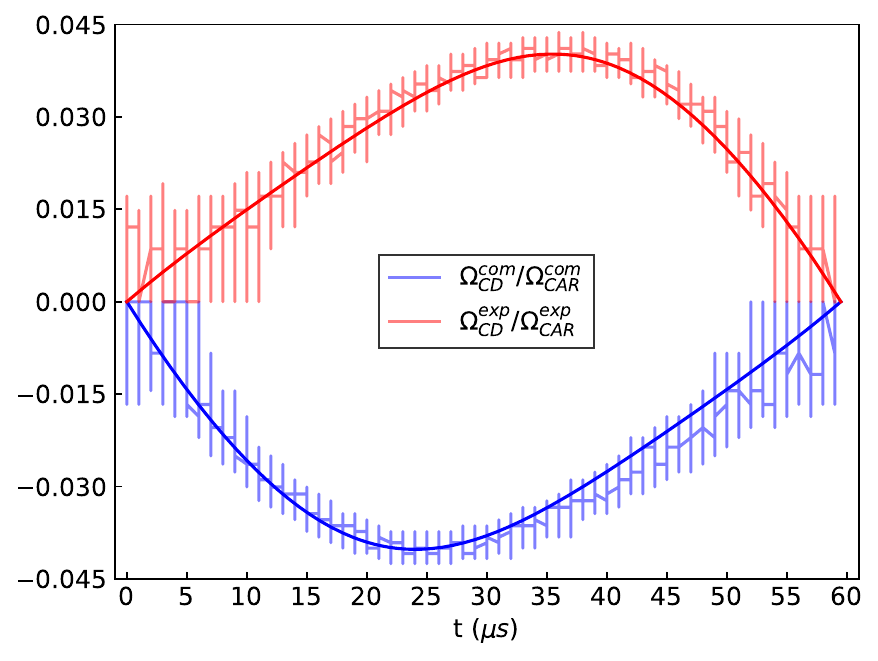}
	\caption{Measured amplitude proportion of CD laser to carrier laser during the compression and expansion strokes, with $\Omega=2\pi\times0.159\ \rm{MHz}$, $\omega_z=2\pi\times2.0338\ \rm{MHz}$, $v_0=\omega=2\pi\times0.075\ \rm{MHz}$ and $\tau=119\ \rm{\mu{s}}$. ${\Omega}_{CD}^{com}$ is the measured amplitude of CD laser at different times $t$ during the compression stroke, while ${\Omega}_{CAR}^{com}$ is the measured amplitude of carrier laser at the same time $t$. The measured results during expansion stroke are denoted as ${\Omega}_{CD}^{exp}/{\Omega}_{CAR}^{exp}$ correspondingly. The translucent lines with noise are the results of sampling measurements from an oscilloscope. Solid lines represent the corresponding numerical simulations.}
	\label{fig:figureS3_CDlaser}
	\end{center}
\end{figure} 

\begin{figure*}[t!]
	\begin{center}
	\includegraphics[width=2.1\columnwidth]{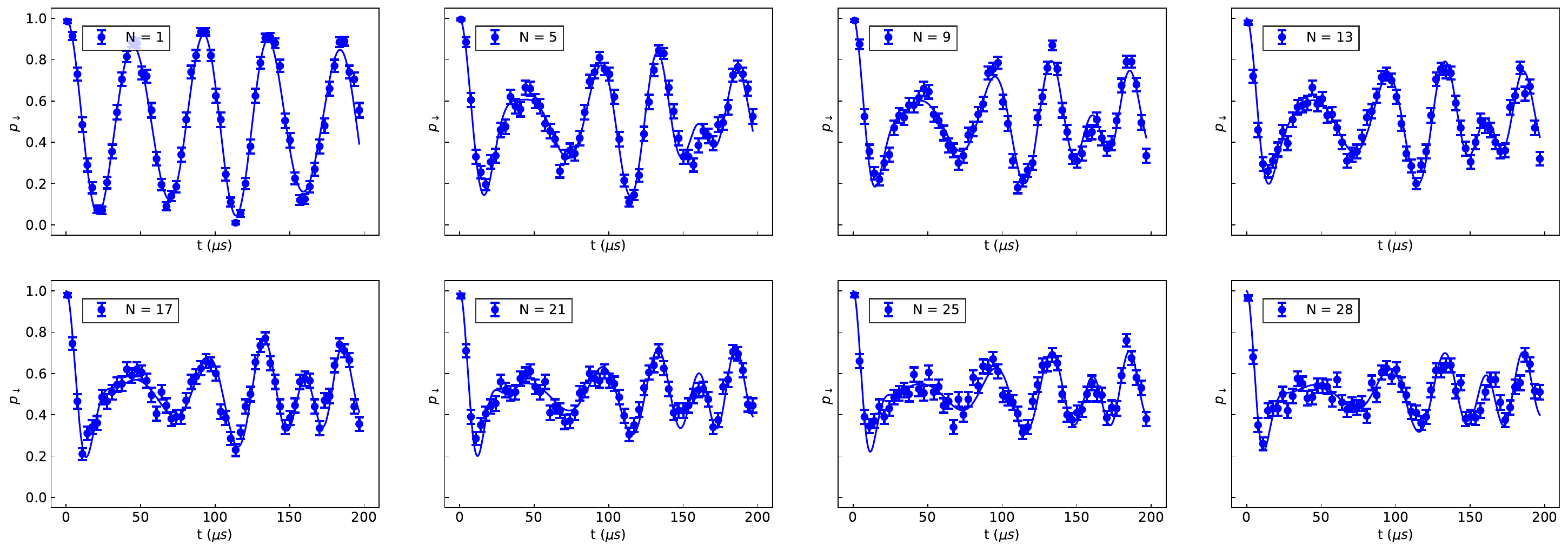}
	\caption{Sample data for the measurement of $p_{\downarrow}$ state population at time $t$ after different number of successive Otto cycles $N$, with $\Omega=2\pi\times0.159\ \rm{MHz}$, $\omega_z=2\pi\times2.0338\ \rm{MHz}$, $v_0=\omega=2\pi\times0.075\ \rm{MHz}$ and $\tau=119\ \rm{\mu{s}}$. Blue circles are the measured average $p_{\downarrow}$ state population of the engine with STA scheme. Error bars correspond to 1 standard deviation of the normal distribution from the raw data which samples 200 times. Solid lines represent the corresponding numerical simulations.}
	\label{fig:figureS4_bsb}
	\end{center}
\end{figure*} 

During the expansion and compression strokes of one Otto cycle, a laser beam with wavelength of $729\ \rm{nm}$ is used to drive the resonant transition between the spin states $\{\ket{\uparrow},\ket{\downarrow}\}$, and to drive the coupling of the spin to phonons, where the relative spin-phonon coupling strength is decided by $\eta\Omega$. In the experiment, we apply an acousto-optic modulator (AOM) connected to power-amplified signal sources to induce laser field to the ion, where the signal is sourced by an arbitrary waveform generator (AWG) \cite{Phys.Rev.Lett.129.250501}. By programming the AWG with desired waveforms, we generate the special three-color light field that can perform frequency scanning and intensity modulation at the same time, as shown in Fig.~\ref{fig:figureS1_laser}. When it is needed to introduce the STA techniques, we would add another carrier component to the three-color light field with a $-\pi/2$ phase difference to ${H_E}(0)=\frac{\Omega}{2}\sigma_x$, as described in Eq.~\eqref{Eq:Ham_9}, whose intensity and frequency are also programmed by AWG.

To make the analysis of the STA cost with the improvement it brings, we measured the amplitude of the CD laser during compression and expansion strokes by detecting its laser intensity onto a photo diode (PD). With a clock-synchronized oscilloscope, we are able to collect and analyze the intensity of the CD laser at different times in one Otto cycle. The intensity of the carrier laser at the same time is also measured through PD and oscilloscope. As shown in Fig.~\ref{fig:figureS3_CDlaser}, the sampling interval of the oscilloscope is $1\ \rm{\mu{s}}$, and the number of samples per point is 2000. By dividing the amplitude of the CD laser by the corresponding amplitude of the carrier laser, we calculated that the time-averaged integration value of this proportion only takes $2.6(2)\%$, while the corresponding numerical simulation result of this value is $2.5\%$.

\subsection{Measurement of the average phonon number}
The experiment begins with the $397\ \rm{nm}$ laser in conjunction with a repumper laser at $866\ \rm{nm}$ to Doppler cool the ion motion, then a series of controlled $729\ \rm{nm}$ laser and $854\ \rm{nm}$ repumper laser pulses are used to sideband cool the ion motion to the ground state $\ket{0}$ and initialize the spin state to $\ket{\downarrow}$ \cite{Phys.Rev.Lett.83.4713}. We then apply a sequence of $729\ \rm{nm}$ laser pulses in conjunction with $854\ \rm{nm}$ repumper laser to drive several successive Otto cycles. We stop the engine evolution after a certain number of Otto cycles, and measure the average phonon number, which corresponds to the work produced by the heat engine on the external system. To measure the average phonon number, we initialize the spin state to $\ket{\downarrow}$ state with negligible effect on the phonon state, then we drive the blue sideband transition between $\ket{\downarrow,n}$ and $\ket{\uparrow,n+1}$ for various time interval $t$. The interaction Hamiltonian of the blue sideband (bsb) transition is:
\begin{equation}
    H_{int}^{bsb}=\frac{\Omega^{bsb}}{2}(e^{-i({\delta}t+\phi_L)}\sigma_{+}\exp\{i{\eta}(a{e}^{-i\omega_{z}t}+{\textit{a}^\dag}{e}^{i\omega_{z}t})\}+h.c.)
\label{Eq:Ham_10}
\end{equation}
{\noindent}where $\delta=\omega_L-\omega_0=\omega_{z}$. In this interaction picture, the wavefunction of the tensor space consisting of spin and momentum can be written as $\Phi(t)=\sum_{m_{z}=\uparrow,\downarrow}\sum_{n=0}^{\infty}{C}_{m_{z},n}(t)\ket{m_{z},n}$, where $\ket{m_{z}}$ and $\ket{n}$ are the time-independent spin and motion eigenstates. If the transitions are coherently driven between $\ket{\downarrow,n}$ and $\ket{\uparrow,n+1}$, the coefficients ${C}_{m_{z},n}(t)$ are given by Schr\"{o}dinger’s equation $i\partial\Phi/\partial{t}=H_{int}^{bsb}\Phi$ as \cite{J.Res.NIST.103.259–328}:
\begin{gather}
\Dot{{C}_{\uparrow,n+1}}=e^{i\phi_L}\Omega_{n+1,n}^{bsb}{C}_{\downarrow,n},\\
\Dot{{C}_{\downarrow,n}}=e^{-i\phi_L}\Omega_{n+1,n}^{bsb}{C}_{\uparrow,n+1},
\end{gather}
\label{Eq:Ham_11}
{\noindent}where $\Omega_{n+1,n}^{bsb}=\Omega^{bsb}|\bra{n+1}e^{i\eta(a+{\textit{a}^\dag})}\ket{n}|$. For an initial state of $\Phi(0)=\sum_{n=0}^{\infty}{C}_{\downarrow,n}(0)\ket{\downarrow,n}$, the coefficients ${C}_{m_{z},n}(t)$ during evolution can be solved as:
\begin{equation}
\begin{split}
    \Phi(t)=\sum_{n=0}^{\infty}{C}_{\downarrow,n}(0)&(\cos(\Omega_{n+1,n}^{bsb}t)\ket{\downarrow,n}\\
    &+e^{i\phi_L}\sin(\Omega_{n+1,n}^{bsb}t)\ket{\uparrow,n+1})
\end{split}
\label{Eq:Ham_12}
\end{equation}
with
\begin{equation}
p_{\downarrow}(t)=\sum_{n=0}^{\infty}p_{n}\cos^{2}(\Omega_{n+1,n}^{bsb}t)
\label{Eq:Ham_13}
\end{equation}
{\noindent}Thus by fitting the resultant $\ket{\downarrow}$ state population, we can reconstruct the population $p_{n}$ of different phonon eigenstates $\ket{n}$. Here, Eq.~\eqref{Eq:Ham_13} is used as the fitting function to obtain the possible results of phonon population distribution $p(n)$ through curve fitting. When estimating the error bar of the fitting results, we make a common assumption of independent and identically distributed Gaussian noise of the experimental data, and the fitted parameters also follow a joint Gaussian distribution. In Fig.~\ref{fig:figureS4_bsb}, we use the data collected in the multiple-cycle quantum heat engine experiment with STA scheme as an example to illustrate our measurement results of $p_{\downarrow}(t)$, which correspond to the $\Bar{n}_{STA}(N)$ results as shown in Fig.~2 in the main text.

\begin{figure}[t!]
	\begin{center}
	\includegraphics[width=1\columnwidth]{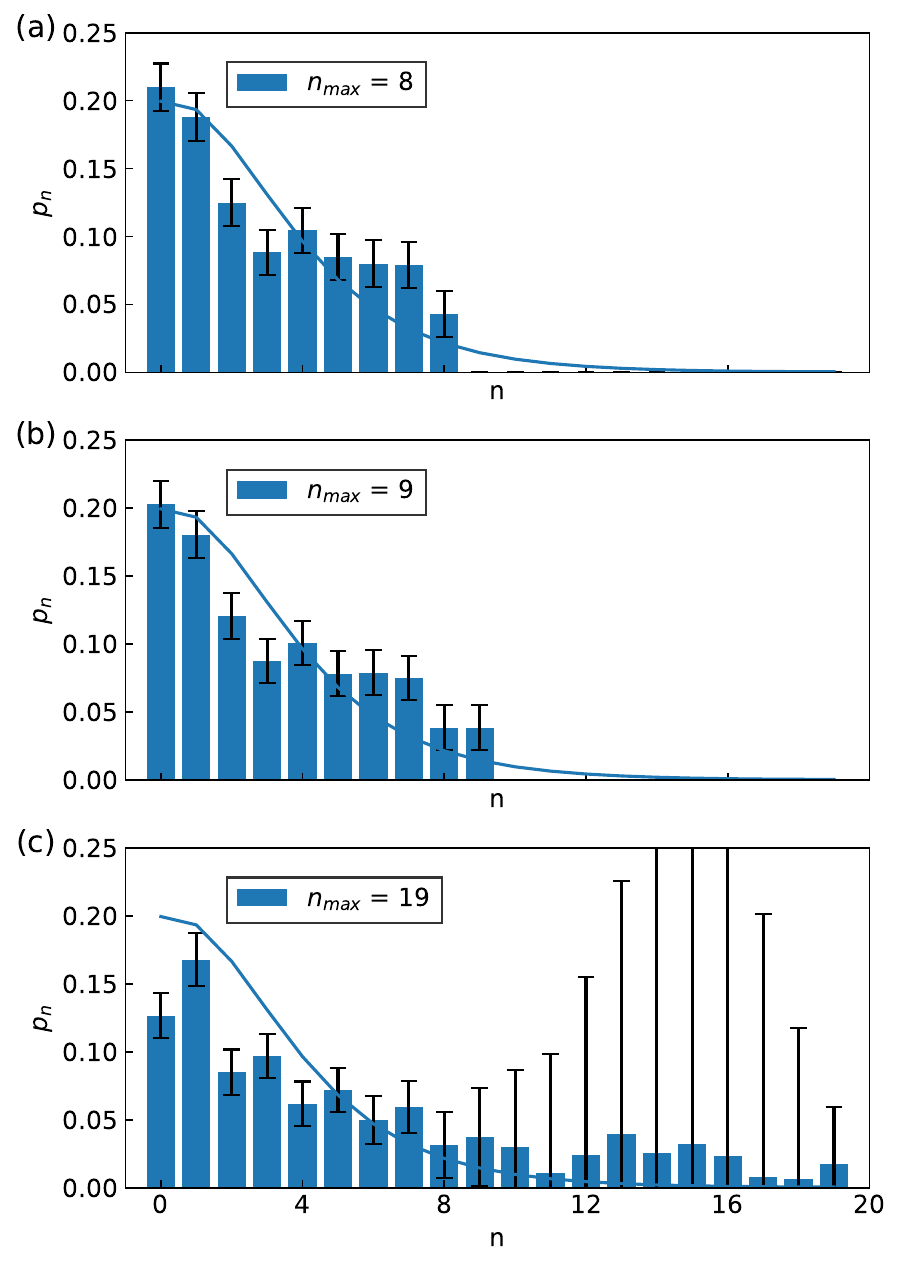}
	\caption{Phonon number distribution $p_{n}$ with different cutoff number $n_{max}$. (a) Phonon number distribution with cut-off number $n_{max}=8$, the extracted average phonon number is $2.9(2)$ while the total occupation $\sum_{n=0}^{n_{max}}p(n)$ is above $95.5\%$. (b) Phonon number distribution with $n_{max}=9$, the extracted average phonon number is $3.1(3)$ and $\sum_{n=0}^{n_{max}}p(n)>96.5\%$. (c) Phonon number distribution with $n_{max}=19$, the extracted average phonon number is $6(8)$ and $\sum_{n=0}^{n_{max}}p(n)>99.5\%$. The bar charts are the results of curve fitting. Error bars correspond to 1 standard fitting error of the phonon state population. Solid lines represent the corresponding numerical simulations.}
	\label{fig:figureS5_nmax}
	\end{center}
\end{figure} 

In the phonon number distribution fitting, in order to avoid the occurrence of overfitting, we use the lowest cut-off number $n_{max}$ and ensure that the total occupation of all the phonon eigenstates is above $95\%$, that is $\sum_{n=0}^{n_{max}}p(n)\geq0.95$ \cite{Nat.Commun.12.1126}. Our principle of choosing the proper $n_{max}$ is to make the fitting error bar of the average phonon number smaller, while the total occupation still exceeds $95\%$. Here, we take the phonon number distribution of the $\Bar{n}_{STA}$ result at $N=28$, which has the largest average phonon number in this experiment, as an example to show how we choose a proper $n_{max}$. As shown in Fig.~\ref{fig:figureS5_nmax}(a), according to the numerical simulations, for this phonon state, when the cut-off number reaches $n_{max}=8$, the total occupation $\sum_{n=0}^{n_{max}}p(n)$ starts to be above $95.5\%$. In the phonon number distribution fitting, the extracted average phonon number is $2.92$, while the fitting error of the average phonon number is $0.24$. When we continue to increase the cut-off number to $n_{max}=9$, with the total occupation is around $96.5\%$, the results of the phonon number distribution are similar to the results of $n_{max}=8$, and the average phonon number is $3.14(0.28)$, as shown in Fig.~\ref{fig:figureS5_nmax}(b). However, if we want to further increase the total occupation to be above $99.5\%$, since for this phonon state, its high-phonon population only accounts for a small proportion of the total occupation, we need to increase the cut-off number to $n_{max}=19$. As shown in Fig.~\ref{fig:figureS5_nmax}(c), the error bar of the state occupation after $n=8$ becomes dramatically large, which results in the curve fitting result of the average phonon number to be $5.6(8.3)$, indicating that overfitting occurs. Hence, we chose the cut-off number as $n_{max}=8$. In order to ensure that the average phonon number extracting method is consistent for all the experimental data at different average phonon numbers, we uniformly set our principle of choosing the proper $n_{max}$ to be that make sure the total occupation exceeds $95\%$.

On the basis of choosing a proper cut-off number $n_{max}$, another method that can be added to further reduce the fitting error is to give the function form of the phonon number distribution $p_{n}=f(n)$, thereby reducing the number of fitting parameters and avoiding the occurrence of overfitting. For example, an ion’s motional state under background heating is given by a thermal distribution $p_{n}={(\frac{\Bar{n}}{\Bar{n}+1})}^{n}$ with the average phonon number of $\Bar{n}$. In this experiment, it can be proved that the distribution of work driven by the Hamiltonian $H_B+{H_{EB}}(t)$ [Eq.~\eqref{Eq:Ham_1}] can be decomposed into a series of superpositions of exponential functions, and the relative coefficients of different series are determined by the initial state \cite{Phys.Rev.E.78.011115}. Therefore, for the experimental results that can already be proved in the simulations that its high-phonon population distribution has the form of an exponential function, we uniformly assume that it has the function form of $p_{n>{n}_{0}}=Aexp(-B(n-{n}_{0}))+C$, where $0<{n}_{0}<n_{max}$.

The reason why the total occupation only exceeds $95\%$ may be due to the state preparation and measurement error. In the experiment, the motional heating rate in our setup is around $240\ \rm{phonon/s}$, which results in the motional decoherence effect. Other sources of errors can be from the fluctuation of the trap frequency $\omega_z$, spin population detection error, and from the phonon number fitting since some noise in the signals may be incorrectly recognized as a high-phonon population and cause the fitting error.

\section{Effect of the shortcut}
\begin{figure}[t!]
	\begin{center}
	\includegraphics[width=1\columnwidth]{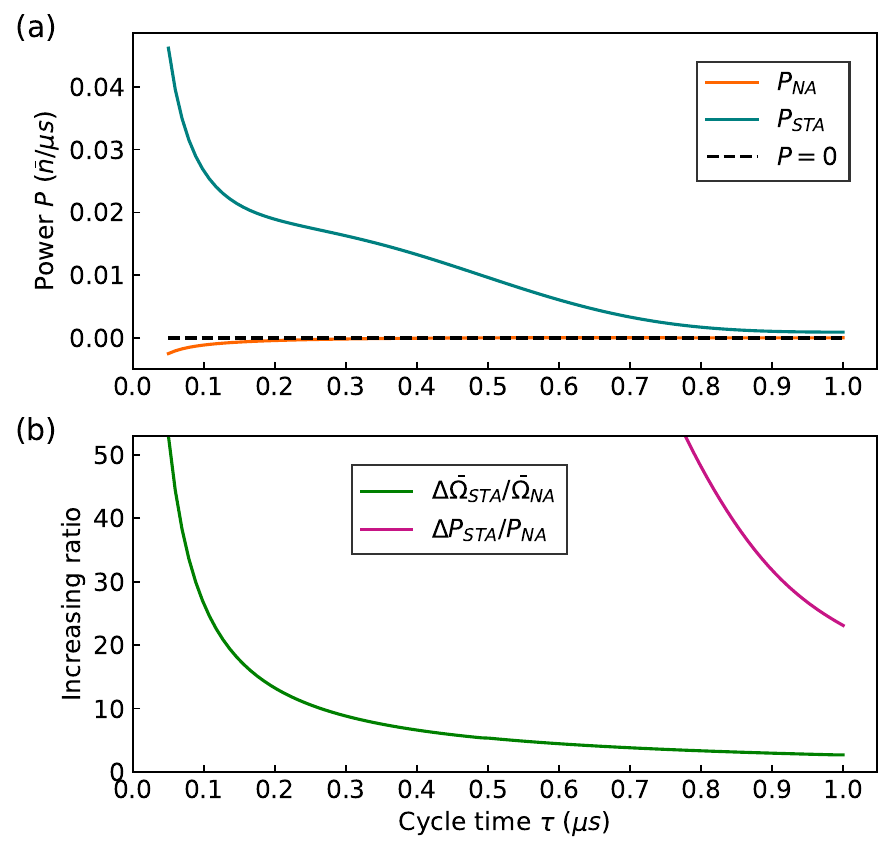}
	\caption{Simulated engine power $P$ and increasing ratio of STA techniques at different cycle time $\tau$, with $\Omega=2\pi\times0.159\ \rm{MHz}$, $\omega_z=2\pi\times2.0423\ \rm{MHz}$, $v_0=\omega=2\pi\times0.075\ \rm{MHz}$ and $N=1$. (a) Orange(Teal) solid line $P_{NA}$($P_{STA}$) represents the numerical simulation results of engine power of the non-adiabatic(STA) engine. The black dashed line represents the case where the output power of the heat engine is zero. (b) Pink(Green) solid line $\Delta \Bar P_{STA}/ \Bar{P}_{NA}$($\Delta \Bar \Omega_{STA}/ \Bar{\Omega}_{NA}$) represents the increasing ratio of engine power(laser amplitude).}
	\label{fig:figureS6_short_time}
	\end{center}
\end{figure} 

\begin{figure}[t!]
	\begin{center}
	\includegraphics[width=1\columnwidth]{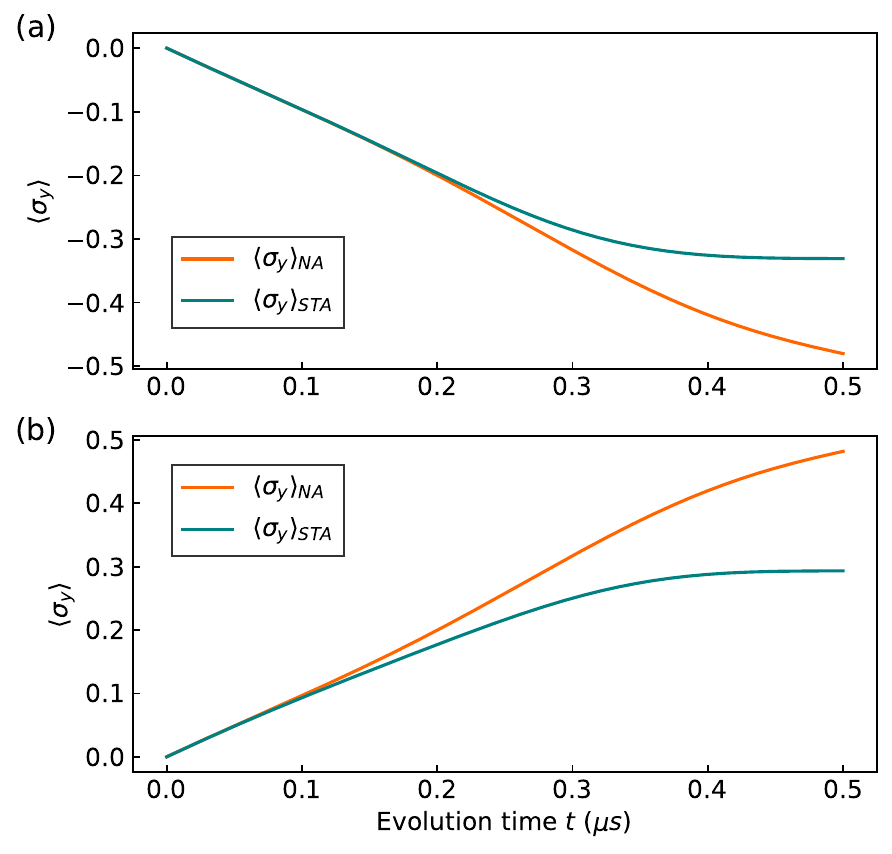}
	\caption{Simulation results of the suppression effect of STA techniques on the expected population along the $\sigma_y$ direction at different evolution time $t$, with cycle time $\tau=1\ \rm{{\mu}s}$, $\Omega=2\pi\times0.159\ \rm{MHz}$, $\omega_z=2\pi\times2.0423\ \rm{MHz}$, $v_0=\omega=2\pi\times0.075\ \rm{MHz}$ and $N=1$. (a) Orange(Teal) solid line ${\langle\sigma_y\rangle}_{NA}$(${\langle\sigma_y\rangle}_{STA}$) represents the numerical simulation results of the expected population along the $\sigma_y$ direction of the non-adiabatic(STA) engine during isentropic expansion. (b) Orange(Teal) solid line ${\langle\sigma_y\rangle}_{NA}$(${\langle\sigma_y\rangle}_{STA}$) represents the numerical simulation results of ${\langle\sigma_y\rangle}$ for non-adiabatic(STA) engine during isentropic compression.}
	\label{fig:figureS7_sigmay_effect}
	\end{center}
\end{figure}

\subsection{Suppression of the irreversible entropy production}
Since non-adiabatic transitions are well-known sources of entropy production that reduce the efficiency of thermal machines \cite{Europhys.Lett.118.40005}, the dynamics of the quantum engine can be sped up with the help of STA techniques to suppress the unwanted non-adiabatic transitions, thereby reducing the associated production of entropy \cite{Phys.Rev.E.98.032121,Phys.Rev.E.99.022110}. The suppression of this irreversible entropy production with the help of STA techniques can be observed when the working medium experiences a fast driving during the work strokes \cite{Phys.Rev.E.99.032108}. Figure~\ref{fig:figureS6_short_time} displays the numerical simulation results of engine power during a fast driving, where the parameters are set the same as Fig.~4 in the main text. We observe in Fig.~\ref{fig:figureS6_short_time}(b) that the output power of the STA engine has a finite value above $P=0$ at very short driving times $\tau<0.1\ \rm{{\mu}s}$, while the non-adiabatic engine is unable to generate work output \cite{Phys.Rev.E.99.032108}.

The reason why we didn't collect and show this kind of experimental results in the main text is that, as mentioned before in the experimental setup, we use the programmed laser pulses to drive the Otto cycles, so due to the limitation of sampling rate and the existence of rising edge from AOMs, it is impossible for us to accomplish such a fast driving in experiments. Besides, the resetting and initial state preparation of the spin takes about $2\sim5\ \rm{{\mu}s}$, which means that for such a fast-driving engine, the heating and cooling strokes will take up most of the time of one cycle, which is also contrary to the common definitions of the Otto cycle. For the above reasons, in our experiments, the total time of one Otto cycle is usually selected to be around $100\ \rm{{\mu}s}$, in which the changing speed is slow so that the evolution of the working medium becomes quasistatic \cite{IVAKHNENKO20231}. Therefore, we can only show the numerical simulation results of this irreversible entropy production phenomenon, but cannot directly demonstrate it in experiments.

\subsection{Suppression of the nonadiabatic transitions} In the experiments, we add a counterdiabatic Hamiltonian $H_{CD}$ to the engine Hamiltonian $H_E$ in order to suppress the nonadiabatic transitions and reduce coherent oscillations along the $\sigma_y$ direction. The suppression of detrimental nonadiabatic transitions with STA techniques can be observed by calculating the expected population along the $\sigma_y$ direction, which is denoted as ${\langle\sigma_y\rangle}$, during the isentropic expansion and isentropic compression of one Otto cycle. Figure~\ref{fig:figureS7_sigmay_effect} displays the numerical simulation results of ${\langle\sigma_y\rangle}$ during the isentropic expansion and isentropic compression of a fast driving with cycle time $\tau=1\ \rm{{\mu}s}$, where the parameters are set the same as Fig.~\ref{fig:figureS6_short_time}. We observe in Fig.~\ref{fig:figureS7_sigmay_effect} that at the start of the engine's expansion and compression process, the initial population along the $\sigma_y$ direction is always 0 since we reset the spin state to the spin-up $\ket{\uparrow}$ (spin-down $\ket{\downarrow}$) state during the heating and cooling process. But with the increasing of the evolution time $t$, the expected population of ${\langle\sigma_y\rangle}$ will gradually become non-zero, resulting in coherent oscillations along the $\sigma_y$ direction. After adding the STA scheme, the absolute value of the expected population along the $\sigma_y$ direction of the STA engine ${\langle\sigma_y\rangle}_{STA}$ is significantly reduced compared with the non-adiabatic engine ${\langle\sigma_y\rangle}_{NA}$, and the gap between them is also widening with the increase of evolution time $t$. Therefore, at the end of the engine's expansion and compression process, the expected population along the $\sigma_y$ direction of the STA engine is only around 0.3, while that of the non-adiabatic engine reaches about 0.5, which proves that the STA scheme can effectively suppress the nonadiabatic transitions and reduce coherent oscillations along the $\sigma_y$ direction.

In the cycles, we perform a sequential coherent operation $U_j$ interleaved by spin projection operation $P_i=\ket{i}\bra{i}$, with $j=\{e,c\}$ represents either expansion or compression and $i=\{\uparrow,\downarrow\}$. The evolution operator after $N$ cycle is:
\begin{equation}
U_N=(U_eP_{\uparrow}U_cP_{\downarrow})^N.
\end{equation}
Since the projection operator is invariant under the rotation along the $z$ axis, i.e. $P_i=e^{i\phi\sigma_z}P_ie^{-i\phi\sigma_z}$ for any $\phi$, one can arbitrarily apply such transformation in between the operators $U_j$ without affecting the results. Thus, the spin coherence is heavily interrupted by each projection. In fact, compared with the case that a desired spin thermal state is prepared between the coherent process, in a single-shot manner the state is indeed coherent however without knowledge with respect to the external drive, which resembles the condition of the above analysis. The motional coherence comes from, instead, the relative phase within $\sigma_y(a+a^\dagger)$, which is intact from the above process, resulting in the build-up of the motional coherence. Thus, we predict the coherence for the spin component at the beginning of each cooling(heating) step does not play a role in the process.



\end{document}